\documentclass[prb,preprint,amsmath,amssymb]{revtex4}
\begin{document}
\author{Kevin Leung$^*$ and Craig Tenney}
\affiliation{Sandia National Laboratories, MS 1415, Albuquerque, NM 87185\\
\tt kleung@sandia.gov (505)8441588}
\date{\today}
\title{Towards First Principles prediction of Voltage Dependences
of Electrolyte/Electrolyte Interfacial Processes in Lithium Ion Batteries}

\input epsf

\begin{abstract}

In lithium ion batteries, Li$^+$ intercalation and processes associated
with passivation of electrodes are governed by
applied voltages, which are in turn associated with free energy changes of
Li$^+$ transfer ($\Delta G_t$) between the solid and liquid phases.
Using {\it ab initio} molecular dynamics (AIMD) and thermodynamic integration
techniques, we compute $\Delta G_t$ for the virtual transfer of a Li$^+$ from
a LiC$_6$ anode slab, with pristine basal planes exposed, to liquid ethylene
carbonate confined in a nanogap.  The onset of delithiation, at $\Delta G_t$=0,
is found to occur on LiC$_6$ anodes with negatively charged basal surfaces.
These negative surface charges are evidently needed to retain Li$^+$ inside the
electrode, and should affect passivation (``SEI'') film formation processes.
Fast electrolyte decomposition is observed at even larger
electron surface densities.  By assigning the experimentally known
voltage (0.1~V vs.~Li$^+$/Li metal) to the predicted delithiation onset,
an absolute potential scale is obtained.  This enables voltage calibrations
in simulation cells used in AIMD studies, and paves the way for future
prediction of voltage dependences in interfacial processes in batteries.

\vspace*{0.5in}
\noindent keywords: lithium ion batteries; voltage prediction;
density functional theory; {\it ab initio} molecule dynamics;
computational electrochemistry

\end{abstract}

\maketitle

\section{Introduction}

Imposing a potential difference between electrodes is the among most basic
operations in electrochemical experiments.  In lithium ion batteries (LIB),
critical processes such as Li$^+$ intercalation into anodes and cathodes,
and electrochemical reductive/oxidative decomposition of organic-solvent
based electrolytes, are governed by half-cell voltages.  For example, widely
used electrolytes based on a mixture of ethylene carbonate (EC), dimethyl
carbonate (DMC, or similar linear carbonates), lithiun ions, and
hexafluorophosphide counterions (PF$_6^-$) start to decompose on the
anode at 0.7-0.8~V relative to lithium metal foil reference (Li$^+$/Li(s)),
while Li$^+$ insertion into commerical graphite anodes occurs at much lower,
0.1-0.2~V, potentials.  To prevent continuous loss of electrolyte and
exfoliation of graphite during charging, anode-passivation by self-limiting
films (called solid-electrolyte interphase or SEI films) formed via
electron-injection-induced electrolyte degradation is critical for LIB
operations.\cite{book2,book,book1,review}  Proposed high-voltage cathode
materials like LiMn$_{1.5}$Ni$_{0.5}$O$_4$ intercalate Li$^+$ above the
experimentally observed stability voltage limit of EC/DMC/LiPF$_6$.  Either
new electrolytes need to be discovered, or the liquid-solid interfaces must
be artificially passivated to avoid electrolyte oxidation on these cathode
surfaces.  To understand and control degradation processes at atomic/electronic
lengthscales, there is arguably an urgent need to use electronic structure
computational tools (e.g., Density Functional Theory, DFT) to calculate the
voltage dependence of interfacial processes.

However, DFT and quantum chemistry techniques deal with fixed numbers of
electrons ($N_e$), not fixed voltages.  The two properties are conjugate,
like pressure and volume; specifying $N_e$ means that potentials are implicitly
defined.  In cluster calculations, where periodic boundary conditions are not
used,\cite{truhlar,dupuis,friesner,batista} intrinsic redox potentials can
be readily calculated at the expense of excluding electrodes in the
models.  {\it Ab initio} molecular dynamics-based redox calculations
with explicit treatment of pure liquid environments have also been an area
of fruitful study.\cite{adriannse} However, calculating potentials on
electrodes in a DFT context and condensed-phase settings has long been
recognized as a challenge in computational
electrochemistry.\cite{halley,sprik10,sprik12,gross08,gross11,otani08,otani06,selloni10,schmickler,santos,arias12,norskov12,neurock06,rossmeisl13,rossmeisl08,rossmeisl11}
It is further complicated by subtle issues of whether ``Galvani''
and ``Volta'' voltages are well-defined, can be measured, or even
computed.\cite{guggen,pratt92,sprik12}

\begin{figure}
\centerline{\hbox{ \epsfxsize=2.50in \epsfbox{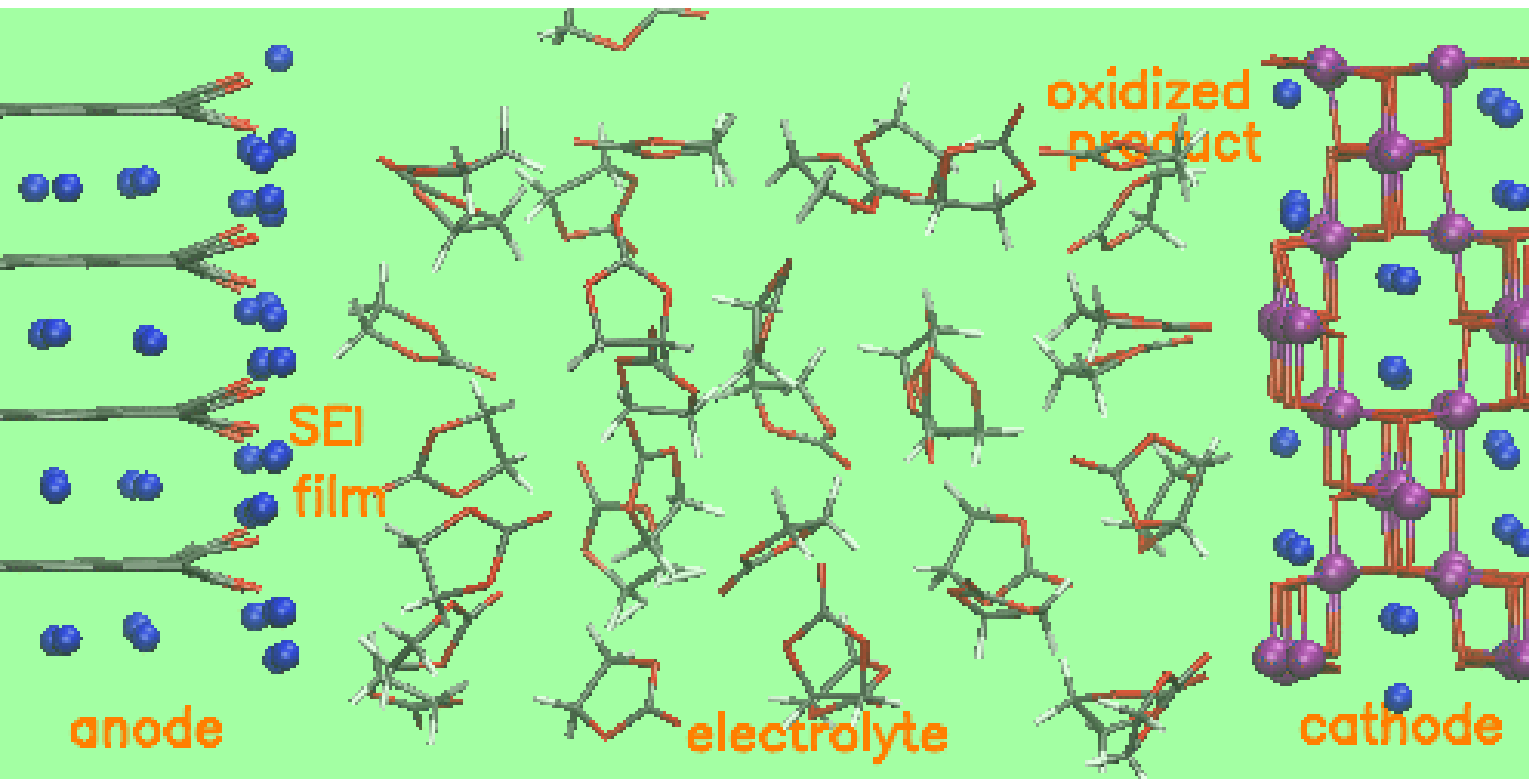} (a)}}
\centerline{\hbox{ \epsfxsize=2.50in \epsfbox{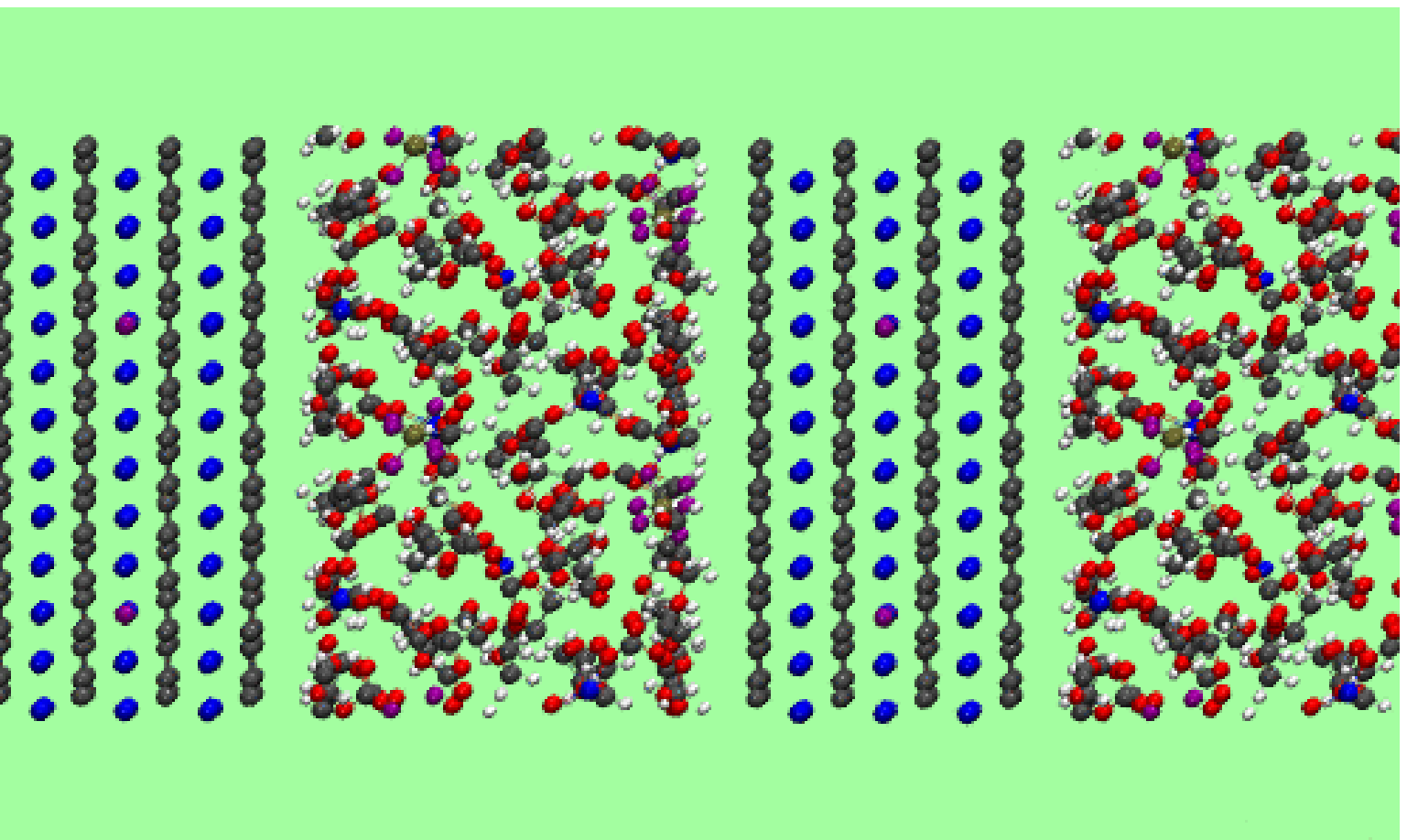} (b)}}
\caption[]
{\label{fig1} \noindent
(a) Schematic of a cathode, an anode, and a liquid electrolyte in the same
simulation cell.  Classical force field electrode models can
predict the voltage drop between two polarizable electrodes, but so far with
not DFT methods.  (b) Periodically replicated simulation cells applied
in this work.  They consist of alternating slabs of LiC$_6$ and liquid EC,
in most cases with Li$^+$, ``Li$^{\lambda+ }$,'' and/or PF$_6^-$ ions solvated
within the liquid region.  The solid and liquid regions are roughly 15 and
19~\AA\, thick, respectively.  Grey, red, white and blue spheres/lines
denote C, O, H, and Li atoms.  
}
\end{figure}

Part of the challenge arises because, unlike classical force fields depiction
of electrodes, e.g., using the polarizable Siepmann-Sprik
model,\cite{siepmann,borodin12a,borodin13,madden,madden1,chandler,pratt_cap}
one-electrode simulation cells are universally applied in DFT
electrochemistry calculations (Fig.~\ref{fig1}).\cite{halley,sprik10,sprik12,gross08,gross11,otani08,otani06,selloni10,schmickler,santos,arias12,norskov12,neurock06} This is partly because accounting for both a cathode and an anode
in a DFT simulation
cell (Fig.~\ref{fig1}a) can easily lead to unphysical $e^-$ migration between
the electrodes.  In lithium ion batteries, the salt concentration is typically
1.0~M, consistent with small a Debye screening length of $\sim$3~\AA, so
anodes and cathodes are indeed well-screened from each other and independent.
But with only one electrode present, imposing a fixed voltage would mean
allowing $N_e$ to fluctuate on the electrode.  This implies permitting
the net charge of periodically replicated simulation
cells to change, which makes the total energy
undefined\cite{saunders,electr,lyndenbell} except in special
cases or with specialized boundary conditions.\cite{pratt96,otani08}
These voltage-specific difficulties are compounded by other
challenges associated with electronic structure modeling of solid-liquid
interfaces,\cite{interface} such as the inherent difficulty of finding
reasonable surface atomic, electronic, and magnetic
structures\cite{meng12,greeley,persson13} and the increased computational
cost of modeling liquids at finite temperature.\cite{pccp,interface}

In this work, we focus on the interface between liquid EC and fully lithiated
graphite (stoichiometry LiC$_6$)\cite{ohzuku} in lithium ion batteries.
When the coulomb efficiency of LIB is close to 100\% (e.g., with well-chosen
voltage windows or passivated electrodes, so that the electrolyte
is no longer being decomposed), the applied voltage should be governed
by Li$^+$ transfer between electrodes and the electrolyte, and
$e^-$ should only move along the external electrical circuit:
\begin{equation}
{\rm Li}_n{\rm C}_{6n} \rightarrow {\rm Li}_{n-1}{\rm C}_{6n}^- + {\rm Li}^+
        {\rm (solv)}. \label{delith}
\end{equation}
The free energy associated with Li$^+$ transfer, denoted $\Delta G_t$
henceforth, is relatively straight-forward, if costly, to compute.  When
$\Delta G_t$ is tuned to zero by adjusting the net surface electronic
density, LiC$_6$ is at the onset of deltihiation -- experimentally known to
occur at 0.1~V vs.~Li$^+$/Li(s).  In other words, the $e^-$ left behind by
delithiation is at a Fermi level ($E_{\rm F}$) 0.1~eV below that of
Li(s) used as reference in batteries.  This fixed point permits concrete
comparison of predictions with measurements.  Further discussions along 
these lines are given in Sec.~\ref{justify}.

$\Delta G_t$ associated with Li$^+$ transfer at liquid EC/solid LiC$_6$
interfaces are conducted using thermodynamic integration (TI) techniques which
closely follow AIMD solvation $\Delta G_{\rm solv}$ calculations.\cite{ion}
Our $\Delta G_t$ calculations are operationally similar to some AIMD pK$_a$
simulations at water-oxide interfaces.\cite{pka1,pka2,pka3}  These methods
fall under the umbrella of ``chemical space'' or alchemical
transformations.\cite{anatole}  The Li$^+$ transfer is virtual; no low
energy, physical pathway exists for Li$^+$ to diffuse from inside LiC$_6$
solid, through the graphite (0001) plane, to the liquid region.  

This paper is organized as follows.  Section~2 provides further justifications
for our approach.  Section~3 describes the thermodynamic integration
method used in bulk liquid EC and at interfaces.  Section~4 discusses
the computed voltages as surface charge densities on electrodes vary, and
compares the predictions with changes in electrostatic potentials and
instantaneous Kohn-Sham band alignments.  A discussion of methodology
improvement is given in Sec.~5, and Sec.~6 concludes the paper.  

\section{Justification of approach}
\label{justify}

\begin{figure}
\centerline{\hbox{ \epsfxsize=2.50in \epsfbox{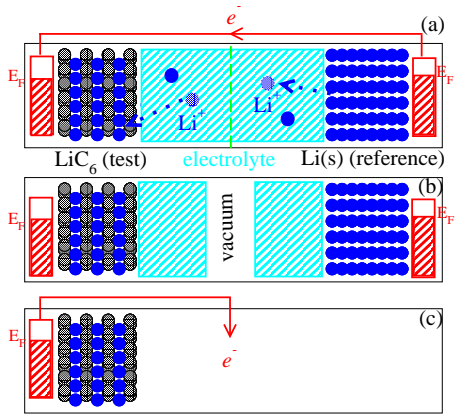} }}
\caption[]
{\label{fig2} \noindent
(a)\&(b) Schematic of working (``test'') and reference electrodes
(Eqs.~\ref{delith} \&~\ref{li_ref}) hypothetically connected by electrolytes
or separated by vacuum, respectively.  Li(s) is never explicitly included
in calculations; only its potential difference with LiC$_6$ is used (see
text).  (c) Charge-neutral LiC$_6$/vacuum interfaces used to illustrate the
computed lithiation potential of zero charge.
}
\end{figure}

A more detailed rationale for our voltage assignment can be made as follows.
The casual reader is encouraged to skip to Sec.~\ref{method}.

The implicit reference electrode is Li$^+$/Li(s) (Fig.~\ref{fig2}a) at its
equilibrium potential:
\begin{equation}
{\rm Li}_n{\rm (s)} \rightarrow {\rm Li}^+ {\rm (solv)} 
			+ {\rm Li}_{n-1}^- {\rm (s)} .  \label{li_ref}
\end{equation}
When an Li$^+$ is removed to the electrolyte, an $e^-$ is left behind at
the metallic Li(s) Fermi level, just as an $e^-$ is left at $E_{\rm F}$
of the working LiC$_6$ electrode (Eq.~\ref{delith}).  If we had used a
two-electrode simulation cell like Fig.~\ref{fig2}a~or~b, the potential of
the working electrode relative to the Li(s) reference can be obtained
without experimental input.  The lithiated graphite potential should
be +0.1 V vs. Li(s), modulo DFT errors.  However, we do not actually
model Li(s) electrodes and their complex, SEI-covered surfaces which exhibit
unknown surface structures and charge densities.  Instead, when both Li(s) and
LiC$_6$ are at their respective $\Delta G_t$=0.0~eV, $E_{\rm F}$ in the two
electrodes differs by 0.10~V according to experiments, and Eq.~\ref{delith}
must be at 0.10~V relative to Eq.~\ref{li_ref}.  In the language of
Ref.~\onlinecite{sprik12}, Eq.~18, Li$^+$ has the same Galvani potentials
inside the electrolyte, LiC$_6$, and Li(s) under these conditions, and the
full-cell voltage is just the difference in $\mu_e$ of the electrodes divided
by $|e|$.

The schematic in Fig.~\ref{fig2}b further clarifies that the ``Fermi levels'' 
of Fig.~\ref{fig2}a can be formally related to measurable work function
differences referenced to the vacuum.\cite{sprik12} Here a vacuum layer
is heuristically inserted in middle of the electrolyte.  The surface
potentials of the two vacuum/liquid interfaces are clearly equal for
the two electrolyte layers covering the electrodes.  They
cancel each other and yield the same potential difference between
the electrodes as Fig.~\ref{fig2}a.  In other words, Li(s) and LiC$_6$ are
in equilibrium with a liquid EC/Li$^+$ electrolyte at the {\it same} absolute
potential relative to vacuum, regardless of whether the vacuum layer exists.
As discussed below, it is computationally advantageous to avoid vacuum regions.

By incorporating the experimentally known 0.1~V voltage difference,
an external ``electrical circuit'' (Fig.~\ref{fig2}a) is heuristically
completed without directly computing $e^-$ orbital levels in or the
cohesive energy of Li(s).  Ambiguities about the measurability
of half-cell potentials\cite{guggen,sprik12,pratt92} should be avoided.  The
assumption that $E_{\rm F}$ on Li metal is at 1.37~V versus vacuum under
Eq.~\ref{li_ref} conditions, which is analogous to the 4.44~V standard
hydrogen electrode reference in aqueous systems, is not used; this
information is subsumed into the 0.1~V potential difference and LiC$_6$
$\Delta G_t$ calculations which incorporate Li$^+$ solvation effects.

By construction, the hidden Li(s) reference is at equilibrium
($\Delta G_t$=0).  But it is a requirement in our scheme to search for
$\Delta G_t$=0 with LiC$_6$.  $\Delta G_t$ is a function of the surface
electron density ($\sigma$) on LiC$_6$.  For the purpose of this computational
paper, at each $\sigma$, we assign $-\Delta G_t/|e|$ as the voltage of the
anode on short timescales, as if LiC$_6$ were a capacitor or an ``ideally
polarizable'' electrode.  This assumes that $\sigma$ is adjusted on time scales
fast compared to Li$^+$ lithiation/delithiation which may alter the voltage.
Indeed, we have frozen all atoms in the LiC$_6$ electrodes in our simulations.
By calculating $\Delta G_t$ over a range of $\sigma$, the condition under
which Eq.~\ref{delith} is at equilibrium ($\Delta G_t=0$) is obtained.
This aspect appears related to Rossmeisl {\it et al.}'s calculations of
H$^+$ at Pt(111)/H$_2$O interfaces, with the difference that there is no
vacuum region in our work.  We also define the ``lithiation potential
of zero charge'' (LPZC) of LiC$_6$ as the $-\Delta G_t/|e|$ value when
$\sigma$=0.  LPZC is not necessarily measurable, but it will allow a check
against an alternate, thermodynamic estimate that assumes liquid-solid
interface effects are minimal on charge-neutral electrodes (Fig.~\ref{fig2}c).  

In reality, at low voltages, anodes should be covered with SEI films
regardless of its material composition.\cite{book2,book,book1,review,harris}
Our calculations on pristine basal planes are meant to reflect the transient
period before SEI starts to form and covers the anode surface; the predicted
voltage dependence is precisely what is needed to understand initial
SEI formation processes.  The LiC$_6$ stochiometry is chosen over
unintercalated graphite (which is more appropriate at higher potentials)
because the onset of LiC$_6$ delithiation (0.1~V vs. Li$^+$/Li(s)) is much
better established\cite{ohzuku} than the onset of lithiation into graphite;
the latter depends on edge site chemistry.
The relatively unreactive graphite basal terminating surfaces are chosen
to slow down parasitic reactions that may occur while $\Delta G_t$ statistics
are being collected.  In the future, LiC$_6$ edge planes, more pertinent to
lithium intercalation, will be considered.  At the same applied voltage, the
surface charge densities of different crystal facets differ, as do those
of pristine electrodes and electrodes covered with SEI.

A brief comparison between our approach/philosophy, and methods used in the
aqueous electrochemistry (especially fuel cell) literature, is next given.
Many computational electrochemistry applications are based on surface science
methods.  Structural optimization is applied at zero temperature; this extends
to the configuration of electrolyte molecules, if present.  A dielectric
approximation is sometimes used to mimic the finite-temperature liquid
environment.  For catalytic metal electrode surfaces, the theoretical voltage
dependence of molecular reactions is calibrated by a combination of Fermi
level and thermodynamic cycle estimates.  While successful for
fuel cell applications, zero temperature-based methods appear to lack the
versatility to deal with ionic processes, such as Li$^+$ intercalation from
solvent into anodes and the parasitic reactions of its counter ion,
PF$_6^-$,\cite{book,oakridge} in batteries.  Indeed, net charges on electrode
and ions in the electrolyte are seldom explicitly included in T=0~K calculations
because ions are insoluble (i.e., they ``salt out'') in crystalized solvents,
and because simulation cells containing material/vacuum interface must be
charge-neutral for their energies to be meaningful.\cite{except,rossmeisl08}
Most previous assignment of Li potential on LIB electrode surfaces, include
our own, apply this imperfect surface science route.\cite{pccp,meng_review} We
will show that LiC$_6$ basal plane/liquid electrolyte interfaces should in
fact be negatively charged.  Note that combining dielectric continuum
approaches with condensed-phase applications of DFT has shown
promise.\cite{otani06,arias12,galli,marzari}  However, in LIB, both solvents
and salts providing the dielectric environment can participate in
electrochemical reactions.  Using an all-atom DFT treatment permits
unconstrained simulations of their reaction mechanisms, and is prefered.

Recently, metal surfaces covered by thin layers of water have been simulated at
finite temperature using AIMD methods.\cite{gross08,gross11,otani08,otani06}
Even though statistical uncertainties can be significant,\cite{gross08}
average work functions can be unambiguously predicted.  The work function
of an electrode covered with a sufficiently thick liquid layer should yield
electronic band alignment needed to elucidate its potential relative to
vacuum, defined as absolute zero energy.\cite{sprik12}  However, adding a
vacuum layer in LIB models significantly increases the simulation cell size,
particularly because symmetric slabs with liquid layers on both electrode
surfaces are desirable to mimic battery systems.  Here, we are most interested
in calculating the voltage dependence of condensed-phase simulation cells used
to study electrolyte decomposition\cite{pccp,ald,oakridge,mno} and Li$^+$
intercalation dynamcis.\cite{abe,borodin12}  Model systems with vacuum gaps
are no longer the {\it same} simulation cells used for studying the
phenomena mentioned above.  In battery experiments, vacuum or vapor regions
do not exist unless an ionic liquid is used as electrolyte.\cite{sullivan}
It appears ideal to avoid artificially opening vacuum gaps in simulations if
an alternative connection to experimental fixed points can be established.

Finally, we compare our approach with rigorous, seminal theoretical studies
of band alignment at TiO$_2$/H$_2$O interfaces.\cite{sprik10,sprik12}
Sprik {\it et al.} reference their TiO$_2$ electronic bands to the standard
hydrogen electrode (SHE) potential at the pH of zero charge of TiO$_2$.
\begin{eqnarray}
{\rm TiOH}^- + {\rm H}^+ {\rm (aq)}  &\rightarrow &
{\rm TiOH}^. + {\rm H}_2  \nonumber \\
{\rm Ti}_2{\rm OH}^+ + {\rm OH}^- {\rm (aq)} + (1/2){\rm H}_2 &\rightarrow &
{\rm Ti}_2{\rm OH}^. + {\rm H}_2{\rm O} {\rm (aq)}.
                \label{she}
\end{eqnarray}
All pocesses associated with Eq.~\ref{she} are computed in water-filled
simulation cell containing the oxide slab.  The pH of zero charge of
TiO$_2$ has been measured, and properties computed at the PZC can
be compared to experiments.  In LIB, potentials of zero charge, determined
not by pH but by the net electron surface density ($\sigma$), are unknown.
Furthermore, the lack of a liquid H$_2$O environment inside LIB renders
calculating Eq.~\ref{she} in LIB simulation cells meaningless.  In LIB
half-cell measurements, lithium metal foils are used as counter electrodes.
Unfortunately, it is impractical to place a Li metal slab in our LiC$_6$
simulation cell and compute reference properties analogous to Eq.~\ref{she}.
Not only is there substantial lattice mismatch, but pristine Li(s) reacts
violently with organic solvents.\cite{ald}  Our approach is agnostic to
the vacuum level or the SHE, but uses the experimental LiC$_6$ voltage
at the onset of Li$^+$ delithiation as reference.  Unlike TiO$_2$, LiC$_6$
is metallic; removing a Li$^+$ from LiC$_6$ leaves an excess $e^-$ on the
Fermi level, not in a localized state.  For LIB cathode oxides which are
electronic insulators, our approach needs to be significantly modified.

\section{Method}
\label{method}

\subsection{Li$^+$ solvation free energy}

The basic thermodynamic integration formula used in this work is
\begin{equation}
\Delta G = \int_\lambda \langle dH(\lambda) /d\lambda
        \rangle_\lambda, \label{eq_solv}
\end{equation}
where $\lambda$ parameterizes the continuous creation or deletion of a
Li$^+$ ion and $\langle \rangle_\lambda$ denotes averaging with intermediate
Hamiltonian, $0<\lambda<1$, sampled using molecular dynamics trajectories.

First we apply it to compute Li$^+$ absolute solvation free energies
($\Delta G_{\rm solv}$) in bulk EC liquid.  Liquid EC simulation cells are
of dimensions (15.2442~\AA)$^3$ and are filled with 32 EC molecules and
a Li$^{\lambda +}$ ion (Fig.~\ref{fig3}a).  Li$^{\lambda+}$ is represented
by a VASP Li$^+$ pseudopotential without core $1s$ electrons and with all
$r$-dependent parts scaled uniformally by $\lambda$.  The success of 
such a scaling has been previously demonstrated in another solvent.\cite{ion}
The integrands $\langle dH(\lambda)/d \lambda \rangle_\lambda$ are
evaluated at either 2 or 6 discrete $\lambda$ points, each sampled every
0.1~ps interval of an AIMD trajectory with a Li$^{\lambda+}$ pseudopotential.
$\lambda$-derivatives are computed via finite difference with $\delta
\lambda$=0.025.  Integrand values at $\lambda$$\neq$$0$ or $1$ are not
physically significant.  They only serve to evaluate the integral in
Eq.~\ref{eq_solv}, which should be path independent.  Other functional forms
for $\lambda$ can be used to scale the pseudopotentials, and should give the
same result.

Since a periodically replicated simulation cell is used in conjunction
with Ewald summation of electrostatics, and cells that contain
Li$^{\lambda+}$ ions have a net $+\lambda$ charge, several electrostatic
corrections are needed.\cite{pratt97,pratt96,saunders,electr,lyndenbell}
The well-known monopole correction is $\alpha$$\lambda^2$/(2$L$$\epsilon_o$),
where $L$ is the box length and $\alpha$ is the Madelung constant.
$\epsilon_o$=1 is imposed for gas phase calculations when evaluating the
energies of bare Li$^{\lambda+}$ which needs to be subtracted while
$\epsilon_o$=$\infty$ is assumed for the high dielectric EC liquid.
The quadrupole correction\cite{ion,pratt92,saunders}
\begin{equation}
\Delta E_{\rm quad} = -2 \eta \pi/3 \int_{\bf r} \rho ({\bf r})
                                ({\bf r}-{\bf r}_o)^2, \label{eq_quad}
\end{equation}
is computed using optimized gas phase EC geometry.  The EC carbonyl carbon,
positioned at ${\bf r}_o$, is chosen as molecular center; $\eta$ and $\rho$
are the EC density and the total (electronic plus nuclear) charge density,
respectively.  Finally, the dipolar contribution to the surface potential,
computed using the carbonyl carbon as molecular center for
consistency,\cite{pratt92} is estimated using classical force fields; see
the supporting information (S.I.) for details.
The corrections arise from the long-range nature of
electrostatics; all higher multiple contributions
vanish.\cite{pratt92,saunders,electr}  

\subsection{Free energy of Li$^+$ transfer from LiC$_6$ to liquid EC}

There is neither sufficient static/dynamic symmetry nor reasonable physical
boundaries to evaluate the quadrupole moment correction (Eq.~\ref{eq_quad})
in simulation cells containing an electrode (Fig.~\ref{fig1}b).\cite{electr}
Consequently, the energies of charged simulation cells require corrections
that cannot be readily evaluated.  In this work, only interfacial cells with
constant $N_e$ and overall charge neutrality are considered.
We freeze all atoms of a ``LiC$_6$'' slab with a C$_{\rm 288}$Li$_{\rm 36}$
stoichiometry (3 layers of Li intercalated between 4 graphite sheet with basal
planes exposed, Fig.~\ref{fig1}b), select one Li in the middle Li slab,
and scale all $r$-dependent parts of its pseudopotential by $(1-\lambda)$.
At the same time, a Li$^{\lambda+}$ pseudopotential is created at a fixed
position in the liquid electrolyte, halfway between electrode surfaces.  Bare
Li$^{\lambda+}$ contributions cancel and their energies do no need to computed.
Eq.~\ref{eq_solv} is approximated with a 2-point gaussian quadrature formula,
which is justified and corrected in Sec.~\ref{solv}.  The simulation cells
have dimensions 34.00$\times$12.96$\times$14.96\AA$^3$.  32~EC molecules,
0~to~4 mobile Li$^+$, and in some cases a Li$^{\lambda +}$ and/or
2~Li$^+$PF$_6^-$ ion pairs fill the gap between electrode surfaces.  The
presence of mobile Li$^+$ mimics experimental conditions, but creates
hurdles for AIMD trajectories which are typically short compared to
Li$^+$ diffusion time scale.  Despite this, we have found reasonable agreement
among predicted $\Delta G_t$ when initial Li$^+$ configurations are varied.

We have only considered removing Li$^+$ from the middle of the LiC$_6$ slab
because these are proof-of-principle ``virtual'' calculations aimed at
elucidating the thermodynamics of Li$^+$ transfer.  In reality, Li$^+$
escapes LiC$_6$ through graphite edge, and Li$^+$ intercalation potentials
at those edge sites often functionalized with oxygen groups will differ
from those inside LiC$_6$.  This will be addressed in future publications.
$\Delta G_t$ variations as a function of Li$^+$ position inside the
{\it electrolyte} is expected to be small, and is discussed at the end of
Sec.~\ref{electrostat}.

The Li$^{(1-\lambda)+}$ and Li$^{\lambda+}$ ions being ``transferred''
are frozen in space.  Hence the TI procedure omits vibrational and
translational entropies of these ions.  Assuming each Li$^+$ is independent
inside LiC$_6$, the Hessian matrix elements $K_{ij}$=$d^2E_{\rm total}
({\bf x})/dx_i dx_j$=3.04, 0.90, and 0.92~eV/\AA$^2$ for $i$=$j$ and are
found to be zero otherwise, where E$_{\rm total}({\bf x})$ is the total energy
of the LiC$_6$ solid and $x_i$ are the Cartesian coordinates of the tagged
Li$^+$.  The Li$^+$ translational/volumetric entropy corresponding to a 1.0~M
salt solution ($\propto$ [$-k_BT$ ln(1660\AA$^3$)]) and its vibrational free
energy ($\propto$ [$-k_{\rm B}T \Sigma_i$ ln($\sqrt{2 \pi k_{\rm B}T/K_{ii}}$])
are added to and subtracted from $\Delta G_t$, respectively.  The correction
amounts to $-0.22$~eV.

\subsection{AIMD Details}

AIMD calculations are conducted using the Vienna Atomic Simulation
Package (VASP) version 4.6\cite{vasp,vasp1} and the PBE functional.\cite{pbe}
A 400~eV planewave energy cutoff, $\Gamma$-point Brillouin zone sampling,
and a 10$^{-6}$~eV convergence criterion are applied at each Born-Oppenheimer
time step, 1~fs in duration.  The $k$-space sampling is spot-checked using a
denser 1$\times$2$\times$2 grid.  The trajectories are kept at an average
temperature of T=450~K using Nose thermostats.
The elevated temperature reflects the need to ``melt'' EC, which has an
experimental freezing point above room temperature, and to improve sampling
efficiency.\cite{water}  In real batteries DMC cosolvent molecules reduce the
viscosity, but DMC is not included herein.  Minor differences in $\Delta G_t$
that may arise from the use of a mixed solvent in real batteries are neglected
in this work.  Tritium masses on EC are substituted for protons.  The first
1~ps of each AIMD trajectory is discarded and the rest is used for sampling
Eq.~\ref{eq_solv}.  The different AIMD simulations and trajectory lengths are 
described in Table~\ref{table2}.

AIMD trajectories are initialized from configurations pre-equilibrated using
simple, rigid-body classical molecular force fields,\cite{pccp,bal} Monte
Carlo (MC) simulations, and the Towhee code.\cite{towhee}   At least 40,000
MC passes at T=1000, 700, and then 400~K are successively conducted, and
the final configuration is used for AIMD simulations.  In that sense,
the electrical double layer should be well-equilibrated to the extent
that the simple classical force field used is accurate.  When PF$_6^-$
and excess mobile Li$^+$ are both present in the electrolyte, the MC
simulation procedure yields Li$^+$/PF$_6^-$, but not 
Li$^{(\lambda +)}$/PF$_6^-$, contact ion pairs.

We have also applied flexible classical molecular force fields to perform
molecular dynamics so as to estimate the dipole contribution to the surface
potential of pure liquid EC.\cite{tenney}  These calculations enable the
prediction of absolute Li$^+$ $\Delta G_{\rm solv}$, defined as the free
energy change of moving an Li$^+$ from vacuum through the liquid-vacuum
interface into EC liquid (see the S.I.).

\section{Results}
\label{results}

\subsection{Li$^+$ solvation in EC liquid}
\label{solv}

\begin{figure}
\centerline{\hbox{ (a) \epsfxsize=2.00in \epsfbox{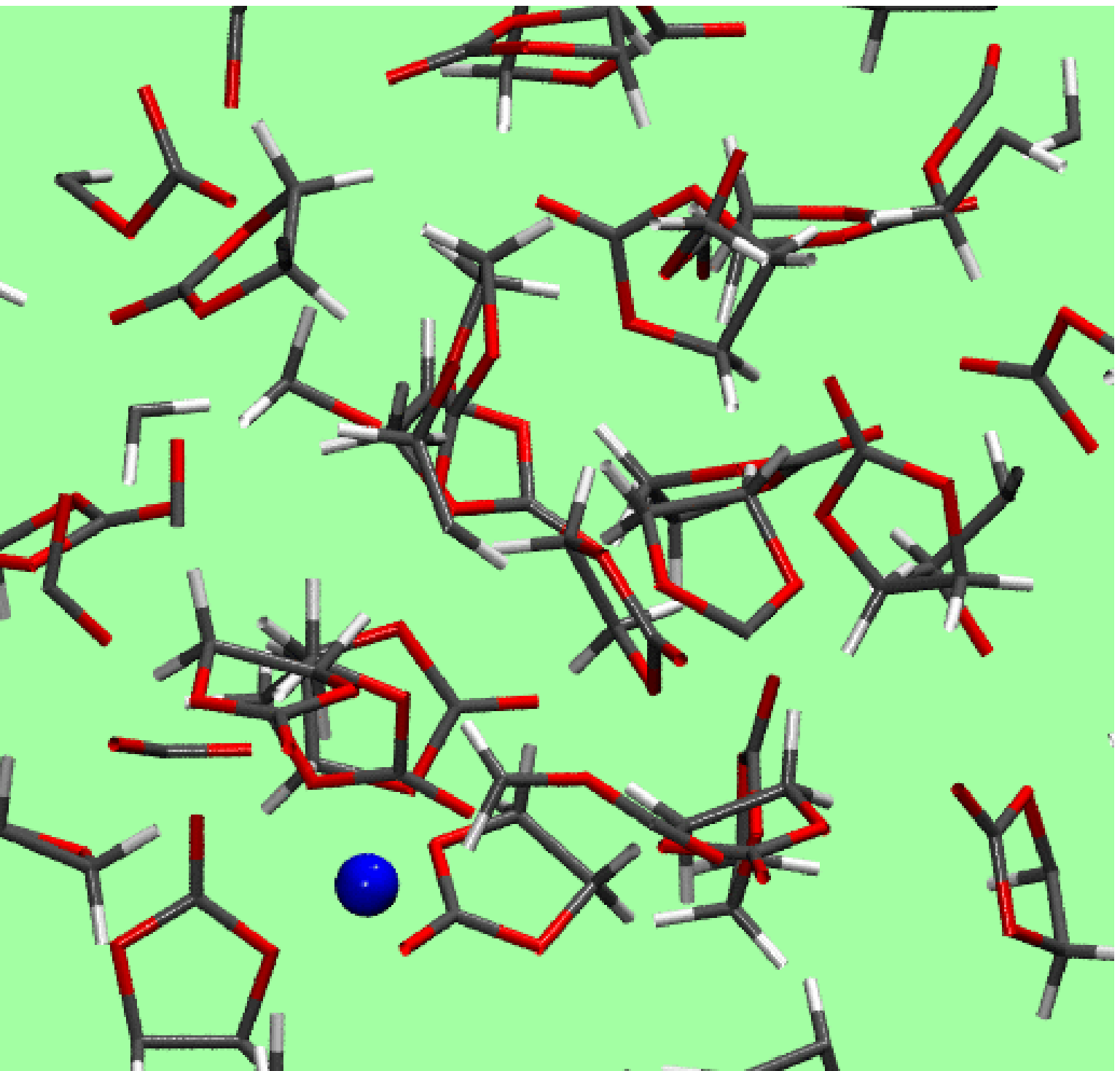} }
            \hbox{ (b) \epsfxsize=2.00in \epsfbox{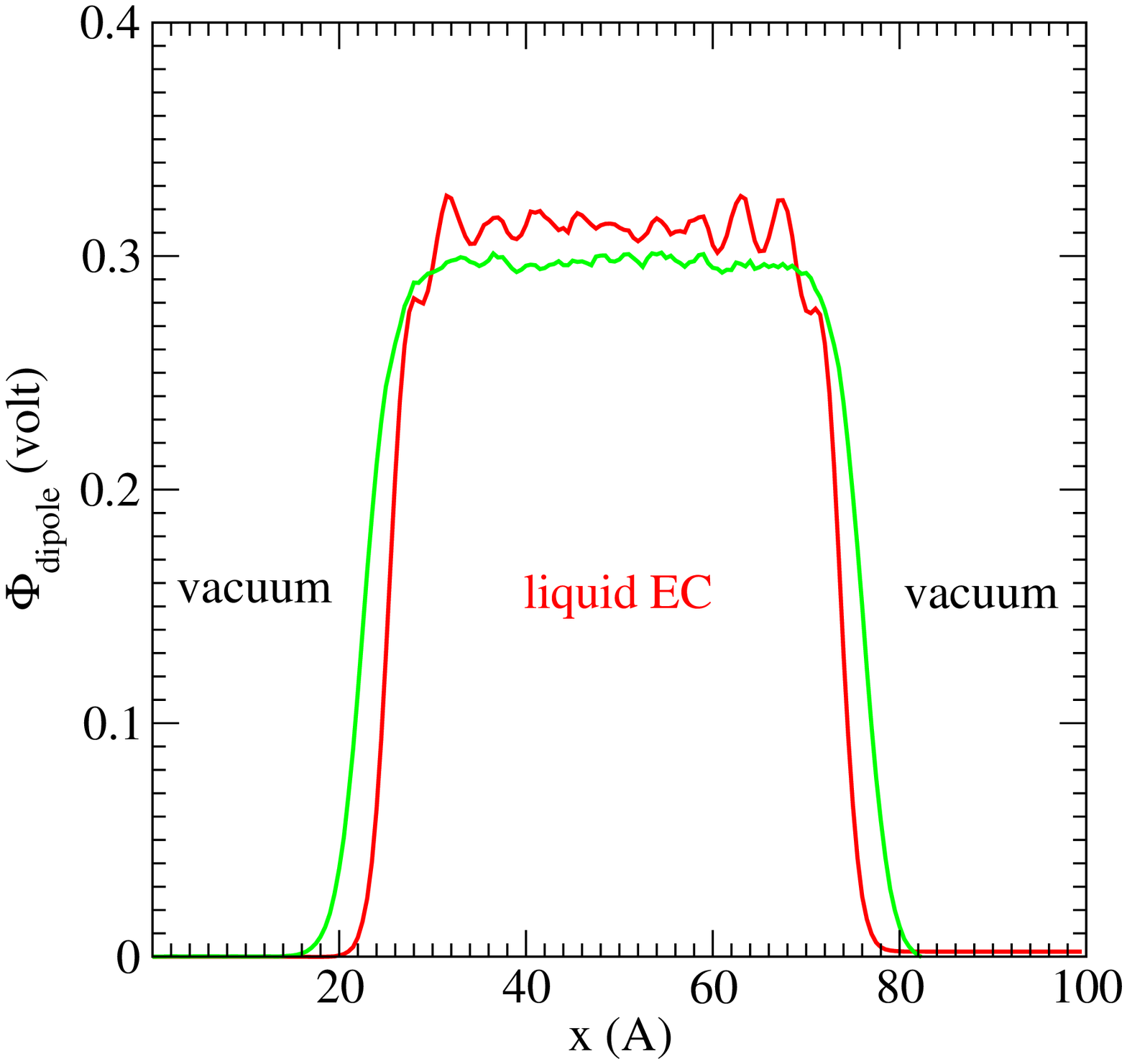} }}
\caption[]
{\label{fig3} \noindent
(a) A snapshot of a Li$^{\lambda +}$ ion in a bulk-like simulation cell.
(b) Dipolar contribution to the surface potential, computed using classical
forcefields at T=350~K (red line) and T=450~K (green).
}
\end{figure}

Although not the main purpose of this work, Li$^+$ 
$\Delta G_{\rm solv}$ calculations illustrate the non-trivial effect of
interfaces and highlights important electrostatic considerations.
Table~\ref{table1} lists $\Delta G_{\rm solv}$ predictions for Li$^+$ in EC
liquid at T=450~K.  Our predicted $\Delta G_{\rm solv}$ is larger in magnitude
than that computed using a bare Li$^+$ plus a dielectric continuum
approximation.\cite{johansson}  It is smaller than $\Delta G_{\rm solv}$
predicted in acetonitrile solvent, reported without surface potential
corrections.\cite{borodin12,liox}  The 2-pt and 6-pt formulas for Li$^+$
solvation differ by 0.15~eV (see the S.I.), which is within twice the standard
deviation, but larger than the discrepancy obtained in H$_2$O
solvent.\cite{ion} In the rest of this work, a global $-0.15$~eV correction
is applied to all $\Delta G_t$ computed using the 2-point formula because
$\Delta G_t$ also involves Li$^+$ solvation effects.  This correction does
not affect the {\it relative} $\Delta G_t$ as the voltage varies.

We have also considered the energy ($E_{C;\lambda}$) of a Li$^{\lambda+}$
embedded in bulk LiC$_6$ solid while all atoms are frozen and $N_e$ is held
fixed.  The S.I. shows that a low-order integration suffices for
$E_{C;\lambda}$.  These findings are used to justify the 2-pt
formula for Eq.~\ref{eq_solv} when we simultaneously annihilate a Li$^+$
inside LiC$_6$ and create a Li$^+$ inside EC liquid.

The dipole contribution to the EC liquid-vapor surface potential is
depicted in Fig.~\ref{fig3}b.  This quantity depends on the choice
of the molecular center; only the sum of the dipolar and quadrupolar
contribution is physical.\cite{pratt92}  Nevertheless, Fig.~\ref{fig3}b
serves to illustrate that the surface potential between a pure liquid and
the vacuum is in general on the order of a fraction of a volt.\cite{pratt87}
When salt is present, some ions may be repelled from the liquid surface while
others may be attracted there, setting up further, non-trivial electric
fields.\cite{jungwirth}  Thus, if we had opened up an artificial gap in the
electrolyte region to estimate band alignment relative to absolute zero
energy, we would have introduced two addition interfaces where distributions
of ions are additional sources of concern.  In our interfacial simulations,
vacuum gaps are avoided, and no classical force field-generated contributions
are included in $\Delta G_t$.

\begin{table}\centering
\begin{tabular}{||c|c|c|c|c||} \hline
$E_{\rm mono}$ & $E_{\rm dip}$ & $E_{\rm quad}$ & 2-pt & 6-pt \\ \hline
2.042 &  +0.300 & +5.389 & -5.054$\pm$0.04 & -5.201$\pm$0.04 \\ \hline
\end{tabular}
\caption[]
{\label{table1} \noindent
Monopole, dipole (dipolar surface potential), and quadrupole correction
to $\Delta G_{\rm solv}$, and the final results, for Li$^+$ solvation in
EC liquid (in eV).  The 6 $\lambda$ integration points
are at 1.0, 0.789. 0.6, 0.4, 0.211, and 0.05 (the last approximates
$\lambda$=0, which cannot be easily computed).  Eq.~\ref{eq_solv} is
evaluated using a trapezoidal rule with these points.  For the 2-point
(Gaussian) quadrature, a 0.5~weight is applied for $\lambda$=0.211 and
0.789.  Statistics for each $\lambda$-point are compiled
over AIMD trajectories of at least 24~ps.  
}
\end{table}

\subsection{$\Delta G_t$ of Li$^+$ transfer from LiC$_6$ to EC liquid}
\label{transfer}

\begin{table}\centering
\begin{tabular}{||l|c|c|c|r|r|r||} \hline
 & $N$(Li) & $N$(PF6$^-$) & $\lambda$ &
 $T$ & $\langle dH(\lambda)/d\lambda \rangle_\lambda$ & $\Delta G_t$ \\ \hline
A&  0 & 0 & 0.211 & 35.7 & +5.32$\pm$0.19 & \\
B&  0 & 0 & 0.789 & 16.7 & -6.57$\pm$0.11 & -1.00 \\ \hline
C&  0 & 0 & 0.211 & 30.1 & +5.52$\pm$0.10 & \\
D&  0 & 0 & 0.789 & 17.9 & -6.54$\pm$0.10 & -0.88 \\ \hline 
E&  1 & 0 & 0.211 & 18.0 & +6.06$\pm$0.14 & \\
F&  1 & 0 & 0.789 & 17.9 & -6.21$\pm$0.05 & -0.45 \\ \hline 
G&  2 & 0 & 0.211 & 15.1 & +6.23$\pm$0.11 & \\
H&  2 & 0 & 0.789 & 18.4 & -5.65$\pm$0.11 & -0.08 \\ \hline
I&  2 & 0 & 0.211 & 18.0 & +6.45$\pm$0.07 & \\
J&  2 & 0 & 0.789 & 16.8 & -5.73$\pm$0.11 & -0.04 \\ \hline 
K&  4 & 2 & 0.211 & 15.6 & +6.58$\pm$0.12 & \\
L&  4 & 2 & 0.789 & 16.0 & -5.82$\pm$0.11 & +0.01 \\ \hline 
M&  4 & 0 & 0.211 &  3.9 & NA & \\
N&  4 & 0 & 0.789 &  3.9 & NA & \\ \hline
O&  0 & 0 & NA & 16.4 & NA & \\
P&  2 & 0 & NA & 15.0 & NA & \\
Q&  4 & 0 & NA &  9.2 & NA & \\ \hline
\end{tabular}
\caption[]
{\label{table2} \noindent
Details of AIMD trajectories.  $N$(Li) and $N$(PF$_6^-$), are respectively
the number of mobile Li$^+$ and PF$_6^-$ ions in the liquid region of the
simulation, and $\lambda$ is the net charge of the frozen Li$^{\lambda +}$
ion if one exists.  The reported trajectory durations ($T$ in picoseconds)
include the first 1~ps equilibration time discarded when collecting statistics.
 Integrands and $\Delta G_t$ are in eV; the latter includes a $-0.22$~eV
entropic correction and a $-0.15$~eV correction for using a
2-point treatment of Li$^+$ solvation (see text).
}
\end{table}

\begin{figure}
\centerline{\hbox{ \epsfxsize=4.50in \epsfbox{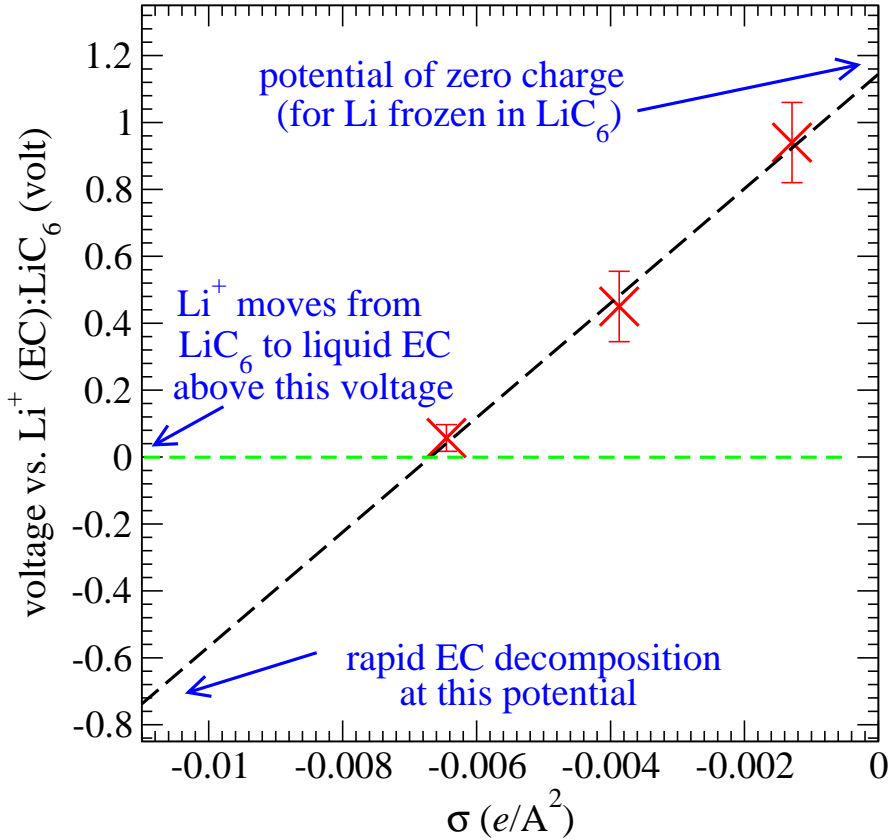} }}
\caption[]
{\label{fig4} \noindent
Predicted potential ($-\Delta G_t$/$|e|$) for virtual Li$^+$ transfer
from the LiC$_6$ slab to the middle of the liquid EC region as the surface
charge ($\sigma$) varies.  Crosses denote the three data points computed,
with 0, 1, and 2 mobile Li$^+$, respectively.  AIMD simulations with 4 mobile
Li$^+$ and no counter ions lead to EC decomposition.
}
\end{figure}

Figure~\ref{fig4} depicts $\Delta G_t$ associated with Li$^+$ transfer
from LiC$_6$ to the EC liquid with 0, 1, and 2 excess $e^-$ in the
LiC$_6$ anode compensated with the same number of mobile Li$^+$ ions in
the electrolyte.  It is compiled using trajectories A-J (Table~\ref{table2}),
and constitutes the main result of the paper.  

The $x$-axis denotes surface $e^-$ density.  In the S.I., we show, as
expected from classical electrostatics, that the excess $e^-$ density
on the electrode is fairly evenly distributed on the two surfaces,
although instantaneously the surfaces can exhibit variations
in charges; the x-axis reflects an average over them (see the S.I.).
Henceforth we report a uniform $e^-$ surface density LiC$_6$ surfaces,
$\sigma=-Q_{\rm Li}/(2A)$, where $A=194$~\AA$^2$ is the lateral surface
area of the simulation cell.  After a Li$^+$ is
transferred from LiC$_6$ to liquid EC, an extra $e^-$ resides on the anode
compared to before Li$^+$ transfer.  In our finite simulation cell,
the resulting change in $\sigma$ is not negligable.  Hence we have marked
the three $\Delta G_t$'s at halfway points, at $\sigma$ values
consistent with $-0.5|e|$, $-1.5|e|$, and $-2.5|e|$ excess $e^-$
in LiC$_6$.  On macroscopic electrodes, adding one extra $e^-$ does not
affect $\sigma$.  By placing our data point at halfway marks, we are
effectively extrapolating towards this infinite size limit. 
The $y$-axis depicts $-\Delta G_t/|e|$ in units of volt.
$y=0.0$~V represents the point where Li$^+$ is equally favored inside
or outside LiC$_6$ anode.  Thus $\Delta G_t$=0 denotes the onset of
delithiation.  Experimentally, this is known to occur at 0.1~V versus
Li$^+$/Li(s).  This reference point allows us to assign an absolute voltage
scale.  Above the green line, LiC$_6$ is thermodynamically unstable.
This emphasizes that anode surfaces, or at least pristine LiC$_6$ basal planes,
need to be negatively charged to retain Li$^+$.

Despite the statistical noise, the three points approximately lie on a
straight line, with $\delta V$/$\delta Q_{\rm e}$$\approx$0.44~V/$|e|$.
Linearity is reasonable over such a small voltage window but is not
essential to our analysis.  If we treat LiC$_6$ as a capacitor, and
factor in the electrode surface area,
$\delta \sigma$/$\delta$(potential)=8.3~$\mu$C/cm$^2$/V.  This value is larger
than the predicted capacitances of unlithiated carbon nanotube arrays in
propylene carbonate\cite{pratt_cap} and porous carbon electrode models in
ionic liquids,\cite{borodin13} and smaller than the capacitances of metallic 
electrodes.\cite{arias12}

\subsection{Statistical Uncertainties and Other Spot Checks}

$\langle dH(\lambda)/d\lambda \rangle_\lambda$ values are tabulated
in Table~\ref{table2}, which also lists statistical uncertainties and
illustrates the dependence on initial configurations.  We have run two
trajectories each for $\lambda$=0.211 and $\lambda$=0.789 with 0 and 2~excess
$e^-$ on the anode.  In all cases, $\langle dH(\lambda)/d\lambda
\rangle_\lambda$ with $\lambda$=0.789 exhibit smaller numerical
uncertainties and dependence on initial
conditions compared to integrands evaluated at $\lambda$=0.211, even when
the AIMD sampling trajectory is longer for the smaller $\lambda$.
We have further considered a third set with two excess $e^-$ on the anode
plus four Li$^+$ and two PF$_6^-$ mobile ions in the electrolyte region
(trajectories K~\&~L).  During equilibration with force fields, each
PF$_6^-$ forms a contact ion pair (CIP) with a Li$^+$.  Such ion-pairing has
been predicted to occur with considerable probability using polarizable
classical force fields.\cite{borodin3}  As might be expected, the
charge-neutral CIPs do not appreciably affect $\langle dH(\lambda)/d\lambda
\rangle_\lambda$ (trajectories K~\&~L).

In the S.I., snapshots of Li$^+$ configuration as well as their distributions
as functions of $x$ are depicted.  They clearly show that Li$^+$ spatial
configurations are not completely converged within 15-35~ps AIMD trajectories.
Despite this, the configurations of large-dipole-moment EC molecules
appear to compensate for differences in Li$^+$ positions and make
$\lambda$=0.789 integrand values relatively consistent (e.g., trajectories
H~and~J).  On the other hand, even $\lambda$=0.211 integrands in simulation
cells {\it without} mobile Li$^+$ (trajectories A~\&~C) exhibit a considerable
dependence on initial configurations.  In contrast, Li$^+$ $\Delta G_{\rm solv}$
calculations conducted in the absence of the electrode does not exhibit 
larger uncertainty at small $\lambda$.  This enhanced sensitivity at small
$\lambda$ is as yet not completely understood.  In Fig.~\ref{fig4}, we have
averaged the results from the two sets of initial conditions and have reported
the error bar as the difference between these two runs.  The exception is the
one mobile Li$^+$ simulation (trajectories E~\&~F); with only one set of
data, we have reported the uncertainty by assuming gaussian distributions of
noise in $\lambda$-windows.  As $\sigma$ becomes more negative, EC orients
themselves so that their partially positively charged C$_2$H$_4$ termini
start to align and point towards the LiC$_6$ surface (not shown).  Detailed
studies of charge-dependent orientational effects are more suited to classical
force field methods\cite{borodin3} than AIMD.

The effect of $k$-point sampling in the lateral ($y$ and $z$) directions
has been spot-checked as follows.  Trajectories G~and~H are sampled every
1~ps.  These configurations are used to calculate $\Delta G_t$ with both 
1$\times$2$\times$2 and $\Gamma$-point Mohhorst-Pack grids.  The resulting
$\Delta G_t$ computed differ by only $-0.04$~eV.  An almost identical small
discrepancy of $-0.04$~eV is found for trajectories A~and~B when using the
two $k$-point grid sizes.  We have not added this small correction to
the final results (Fig.~\ref{fig4}, Table~\ref{table2}).

\subsection{Lithiation Potential of Zero Charge}
\label{pzc}

Extrapolating the results in Fig.~\ref{fig4} to $\sigma$=0 yields the
lithiation potential of zero charge (LPZC) for LiC$_6$ basal planes.  It
is predicted to occur at 1.14~V vs.~Li$^+$/LiC$_6$ (1.24~V vs.~Li$^+$/Li(s)).
At this potential, Li$^+$ should diffuse out of LiC$_6$ into the electrolyte.
Thus LPZC for LiC$_6$ basal planes cannot be measured.  Delithiated graphite
exhibits a potential of zero charge which has no relation to the LiC$_6$
LPZC.\cite{pzc}

A back-of-the-envelope calculation\cite{trasatti} supports the existence
of a positive LPZC.  If interfacial effects and excess negative charge
at basal plane surfaces were absent, the contributions to $\Delta G_t$
can be estimated via the following thermodynamic pathway.  (1) Remove an
Li atom from bulk LiC$_6$; (2) Li(g)$\rightarrow$Li$^+$(g)+$e^-$; (3) put
the ionized $e^-$ back into LiC$_6$ (reverse of the work function,
see the S.I.); (4) Li$^+$(g)$\rightarrow$Li$^+$(EC).
The energies of these processes are computed using DFT/PBE and listed in
Table~\ref{table3}.  They add to a $-1.84$~eV exothermicity for removing
Li$^+$ to the EC liquid, or about $+1.94$~volt versus Li$^+$/Li(s).  An
also identical value of 2.00~volt is obtained by focusing only on the energy
of $e^-$ at the Fermi level, i.e., the LiC$_6$ workfunction, and subtracting
the standard 1.37~V.  This is because the experimental equivalents of
Table~\ref{table3} are present when deriving the 1.37~V Li(s) reference.

In Fig.~\ref{fig4}, $-\Delta G_t/|e|$ is 1.24~V vs.~Li$^+$/Li(s) at $\sigma=0$.
This estimates, and the 1.94~V discussed above, are both large and positive.
Their difference must be due to the neglect of solid-liquid interface effects
in the latter, known to reduce the work function of charge-neutral
water-covered metal surfaces by $\sim 1$~eV.\cite{sprik12,pushback} Other
factors and systematic/statistical errors may also contribute to the difference.
Battery techologists focus on voltage variations and arguably do not have a
pressing need to measure $\sigma$.  However, it is crucial for theorists to
impose the correct explicit surface charge in DFT calculations to represent
realistic, experimental potentials.  This consideration has arguably been
neglected in most AIMD interfacial calculations (see the discussions in
Ref.~\onlinecite{interface}), although the previous works reveal important
electrolyte decomposition mechanisms which should be relevant over
large voltage windows.

\begin{table}\centering
\begin{tabular}{||l|r|r|r|r|r||} \hline
anode & Li in anode & IP (Li gas) & -$\Phi$ (anode)
                & Li$^+$ $\Delta G_{\rm solv}$ &net \\ \hline
LiC$_6$ & +1.65 &  +5.30 & -3.37 & -5.20 & -1.84 \\ \hline
Li(s) & +1.56 &  +5.30 & -3.05 & -5.20 & -1.56 \\ \hline
\end{tabular}
\caption[]
{\label{table3} \noindent
Contributions to $\Delta G_t$ for Li$^+$ transfer from LiC$_6$ or
Li(s) to liquid EC, if liquid-solid interfacial effects were absent, in eV.
They are respectively the binding energy of an Li atom in LiC$_6$ or Li(s),
gas phase Li ionization potential, the negative of LiC$_6$ or Li(s) work
function, and Li$^+$ solvation free energy in EC liquid, all computed using
DFT/PBE.  A $-0.22$~eV entropy correction is added to ``net.''
}
\end{table}

\subsection{Electrostatic/Exchange-Correlation Potentials as $\sigma$ varies}
\label{electrostat}

\begin{figure}
\centerline{\hbox{ \epsfxsize=4.50in \epsfbox{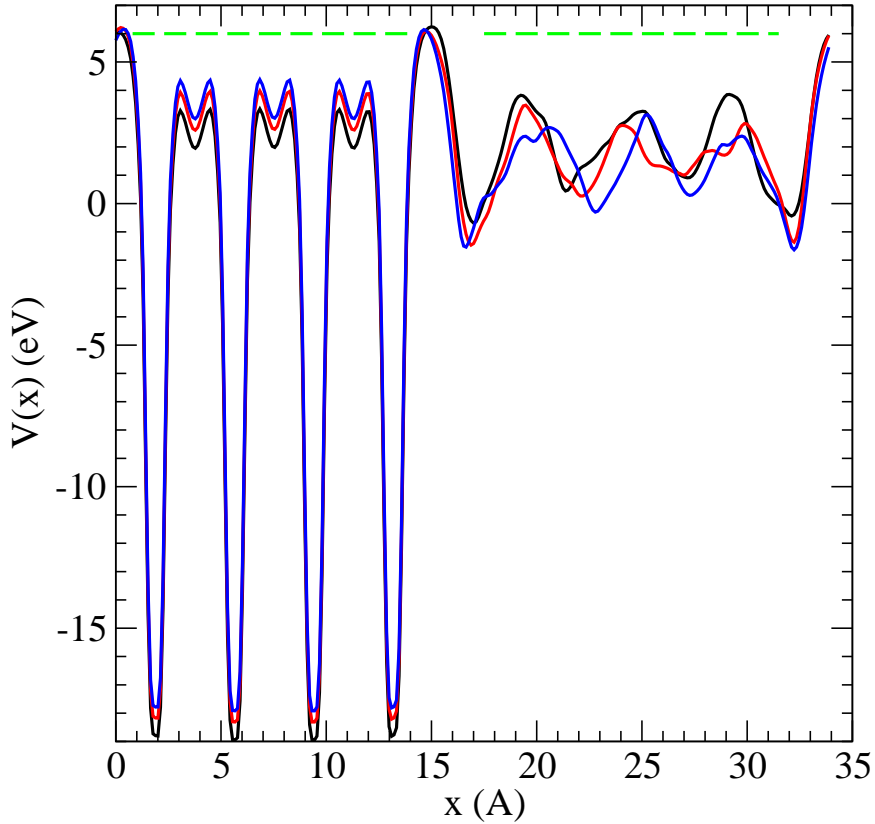} }}
\caption[]
{\label{fig5} \noindent
Average electrostatic-plus-exchange/correlation potential ($V(x)$)
with 0 (black line), 2 (red), and 4 (blue) excess $e^-$ on LiC$_6$.
High $V(x)$ regions repel $e^-$.
}
\end{figure}

Next, we analyze potential differences in cases where 0,~2,~and~4 excess
$e^-$ reside on the LiC$_6$ slab (trajectories O,~P,~and Q) using a
electrostatic analysis complementary to calculating ion transfer free
energies.  The simulation cells considered (O-P in Table~\ref{table2}) do
not contain $\lambda$-scaled lithium ions and are independent of
Li$^+$-transfer $\Delta G_t$ simulations.  Figure~\ref{fig5} depicts the
average potential ($V(x)$) which include electrostatic and DFT
exchange-correlation contributions, sampled every 1.0~ps.  The differences
between $V(x)$ in the electrolyte and LiC$_6$ regions, bracketed by the
green lines in the figure, are $\Delta {\bar V}=$ 5.80, 4.75, and 4.15~V
respectively for 0, 2, and 4~excess~$e^-$.  More excess $e^-$ on the anode
translates into a higher $V(x)$ there (less favorable for electrons to
reside in).  The 1.05~V difference in $\Delta {\bar V}$ between the first
two runs is reasonably similar to the 0.88~V difference observed in
Fig.~\ref{fig4}.  The two values are not expected to be identical because
Fig.~\ref{fig5} averages almost the entire electrolyte region, not at one
value of $x$.  Absolute voltages cannot be estimated from
electrostatic potential differences.

From the similarity in the $\sigma$-dependence of $\Delta G_t$ and
$\Delta {\bar V}$, it may be argued that $\Delta G_t$ only needs to be computed
at one $\sigma$; the $\sigma$-potential relation can then be determined
using $\delta \Delta {\bar V}$/$\delta \sigma$.  This intriguing
alternate strategy comes with the following caveat.  The $\Delta {\bar V}$
prediction for 4~excess~e$^-$ on the anode (Fig.~\ref{fig5}) deviates from
the linear relation of Fig.~\ref{fig4}.  It should be taken with a grain of
salt because Li$^+$ ions adsorbed directly on the basal planes
(Fig.~\ref{fig7}, below) are excluded from the region arbitarily chosen for
$\Delta {\bar V}$ sampling (green lines in Fig.~\ref{fig5}).  Thus
$\Delta {\bar V}$ may depend on how many mobile Li$^+$ ions are included
in the averaging procedure.  The electrolyte is also experiencing
decomposition, and this trajectory has to be prematurely terminated
(see Sec.~\ref{decomp}).

The statistical uncertainties depicted in Fig.~\ref{fig4} can in principle
be measured as time-dependent properties.  The numerical uncertainties
associated with Fig.~\ref{fig5}, which are about 0.15~V, are however
unphysical unless a nanosize probe is used.  Assuming the system size is
well-converged and fluctuations
in each copy of the simulation cell are statistically independent, doubling
the lateral dimensions of the cell (i.e., increasing $A$ by a factor of 4)
should yield a standard deviation half as large.  Such cell size-dependence
fluctuations should be true of previous calculations of work
functions of water-covered metal surfaces as well.\cite{gross08,gross11}

$V(x)$ is rather structured in the nanoconfined electrolyte region,
reflecting solvent layering which is observed even in classical force
field simulations (Fig.~\ref{fig2}b).  We have conducted short AIMD
test runs to show that such layering has little impact on Li$^+$
distributions.  First, a Li$^+$ is frozen at either $x$=23.0~\AA\, or
$x$=24.7~\AA, respectively a valley and peak in the $V(x)$ curve.
The rest of the electrolyte around the fixed Li$^+$ is pre-equilibrated
with classical force fields.  Then AIMD is initiated, with the
tagged Li$^+$ and all electrolyte molecules allowed to move.  In each case,
the tagged Li$^+$ is found to fluctuate in space over the course of a
few picoseconds, but does not exhibit significant net displacement in the
$x$ direction.  The one initially at $x=23$~\AA\, does not fall into
potential wells in $V(x)$ computed in the absence of the tagged
Li$^+$.  This underscores the fact that Li$^+$-solvent interactions are
far stronger than the solvent-solvent interactions which determine $V(x)$.
This observation dovetails with classical force field predictions
that the free energy profile (i.e., ``potential of mean force'') of Li$^+$
displacement towards the electrode is much less structured than the mean
electrostatic potential; it suggests that using Li$^+$ transfer
to calculate electrode voltages may lead to faster convergence with respect
to the thickness of the liquid electrolyte layer.

\subsection{Kohn Sham Band Structure as $\sigma$ Varies}
\label{kohn-sham}

\begin{figure}
\centerline{\hbox{ \epsfxsize=4.50in \epsfbox{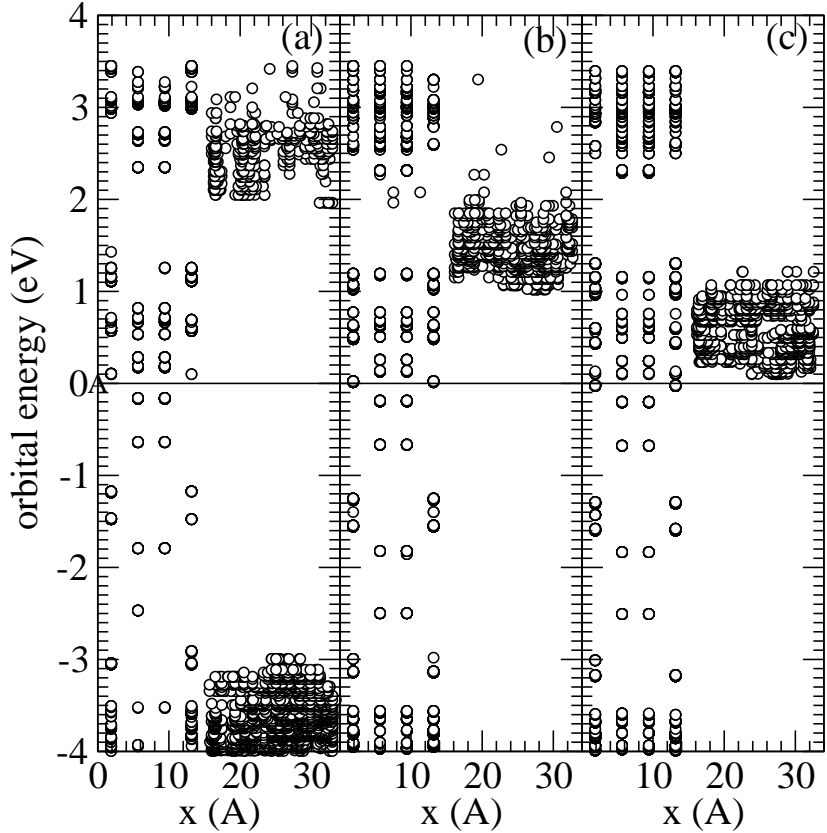} }}
\caption[]
{\label{fig6} \noindent
Instantaneous Kohn-Sham orbital alignment.  The orbital density is decomposed
on to atoms at their $x$-coordinates for (a) 0; (b) 2; and (c) 4 excess $e^-$
on LiC$_6$.  $x<$15~\AA\, denotes the LiC$_6$ region; outside that range
resides the electrolyte.  Fermi levels are at $E$=0.00~eV.  
}
\end{figure}

Next we correlate $\Delta G_t$ with shifts in the bottom of the conduction
band in the electrolyte region as $\sigma$ varies.  Figure~\ref{fig6} depicts
Kohn-Sham band structures in a single snapshot towards the end of trajectories
O,~P,~and~Q, with 0, 2, and 4~excess $e^-$ in LiC$_6$ respectively.
Here the charge density on each Kohn-Sham orbital is split among
atoms $i$ at positions $x_i=x$; if $\rho_i$$\geq$0.005,
the histogram at $x$ is incremented.  The arbitary cutoff means that
some orbitals in the electrode are omitted; the density of state is somewhat
higher than shown in Fig.~\ref{fig6}.  Referenced to $E_{\rm F}$, always
set at 0.0~eV, the electrolyte conduction band minimum is shifted downwards by
roughly 1~eV with each successive injection of 2~excess $e^-$ into the LiC$_6$
slab.  The magnitude of the shift is consistent with both $\Delta G_t$
predictions (Fig.~\ref{fig4}, 0.88~V for every two $e^-$ added) and
electrostatic potential estimates (Fig.~\ref{fig5}, 1.05~V per two $e^-$),
although these shifts may vary somewhat from one snapshot to the next.

Figure~\ref{fig6}c is particular interesting because the snapshot
is taken less than 1~ps prior to an EC absorbing two $e^-$ and
decomposing.  The bottom of the electrolyte conduction band is almost
at the Fermi level, enabling rapid $e^-$ transfer to and decomposition
of the electrolyte.

\subsection{Electrolyte Decomposition at Low Voltages}
\label{decomp}

\begin{figure}
\centerline{\hbox{ (a) \epsfxsize=2.50in \epsfbox{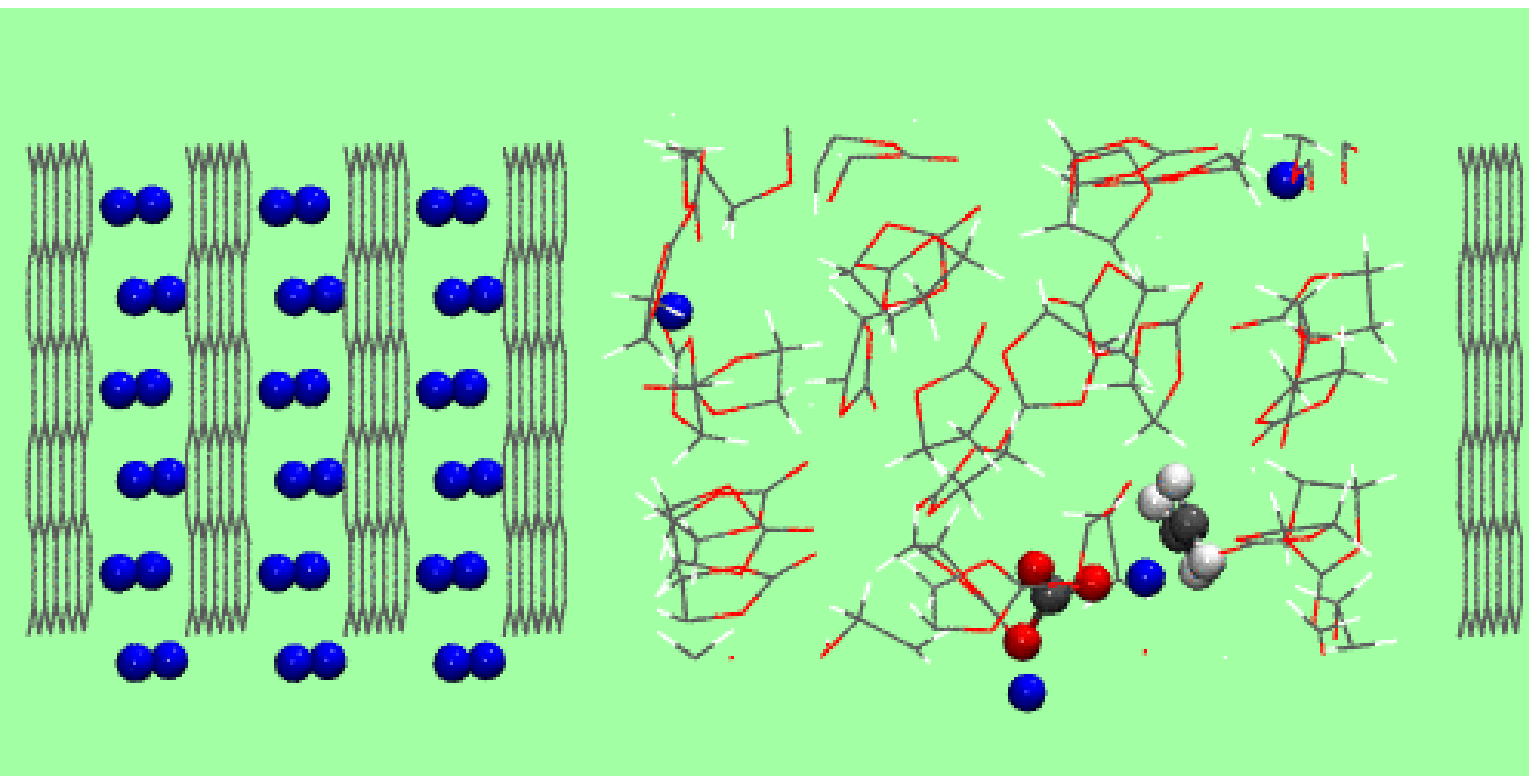} }}
\centerline{\hbox{ (b) \epsfxsize=2.50in \epsfbox{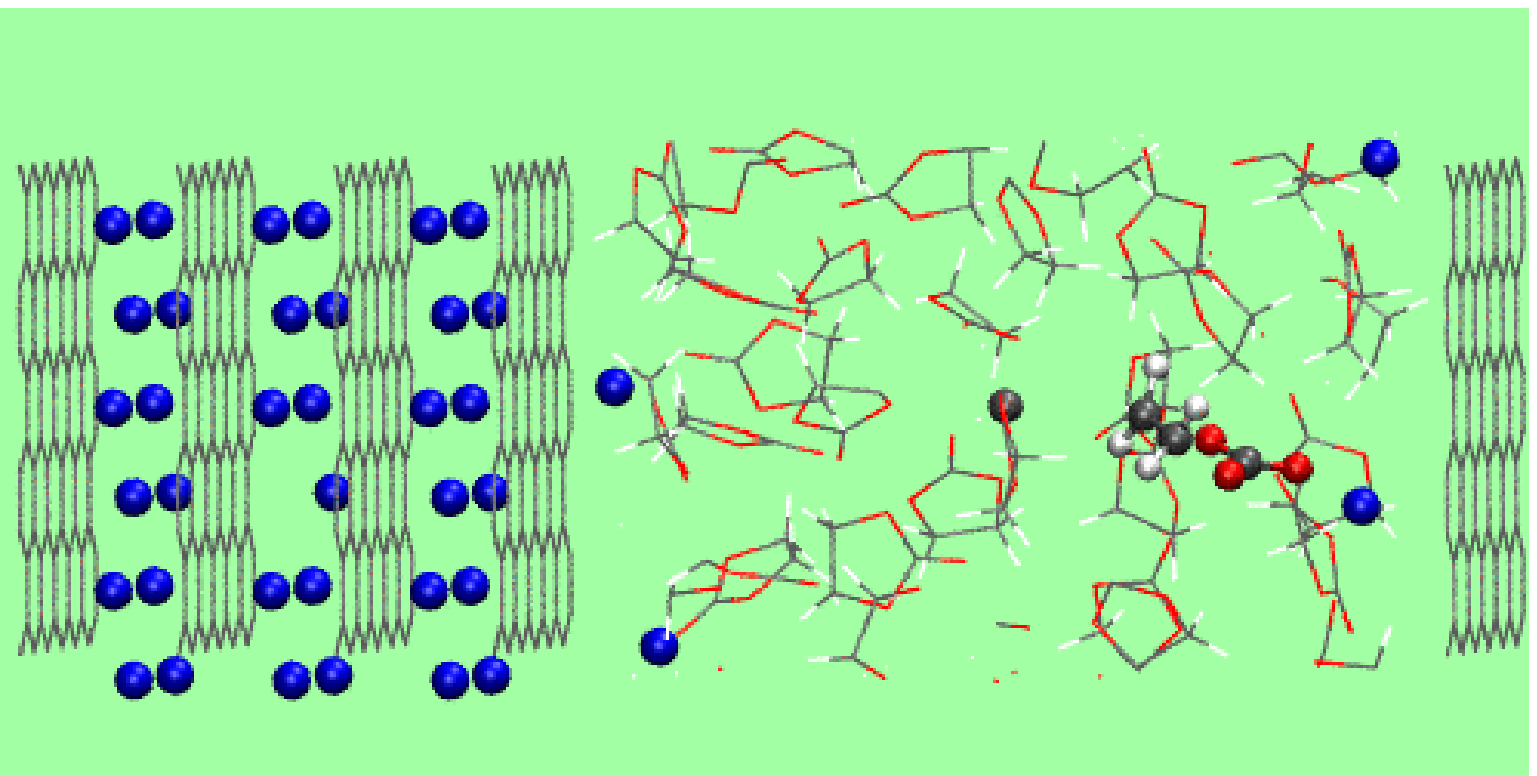} }}
\caption[]
{\label{fig7} \noindent
Two snapshot of trajectories with 4 mobile Li$^+$.  (a) Trajectory Q, after
9.2~ps; (b) trajectory N, after 4~ps (with an extra frozen Li$^{0.8+}$ ion
present).  Decomposed/intact EC molecules are depicted as ball-and-stick/stick
figures, respectively.
}
\end{figure}

AIMD simulations with $>1$~excess $e^-$ in the model anode are metastable.
According to Fig.~\ref{fig4}, their potentials are below the
experimentally known EC reduction potential ($\sim 0.7$~V vs.~Li$^+$/Li(s)).
With 2~excess $e^-$, $-\Delta G_t$/$|e|$ is close to the Li(s) plating
potential ($\sim 0.0$~V vs.~Li$^+$/Li(s)), which is another source of
battery safety concern.  Despite this, electrolyte
decomposition is not yet observed in those short AIMD trajectories because
the basal plane is relatively inert.  However, with at least 4 excess $e^-$ in
the LiC$_6$ slab, a EC molecule absorbs two $e^-$ and decomposes within
picoseconds.  Fig.~\ref{fig7} shows snapshots at the end of trajectories
Q~and~N.  In each case, a EC molecule breaks an oxygen-ethylene carbon
(O-C$_{\rm E}$) bond, not an oxygen-carbonyl carbon (O-C$_{\rm C}$) bond.
Quantum chemistry calculations on EC$^{2-}$ predict that the latter barrier
is smaller.\cite{e2}  However, on a material surface, we have found that
the (O-C$_{\rm E}$) cleavage barrier can be reduced.\cite{ald}  It is also
possible that the extremely low effective voltage associated with these
trajectories has changed the dominant decomposition mechanism.  Given
the limited bond-breaking statistics available in these trajectories,
the potential dependence of bond-breaking mechanism cannot be completely
resolved but is an important consideration for future studies.

The very fast, adiabatic $e^-$ transfer to EC molecules apparently occurs
via fluctuation-induced band-crossing.  As discussed in Sec.~\ref{kohn-sham},
the bottom of the conduction band in the electrolyte region almost
coincides with the Fermi level residing in the electron-conducting anode
(Fig.~\ref{fig6}c).  $e^-$ can thus pour into the liquid region.
This seems to explain why in Fig.~\ref{fig7}a, even an
EC in the middle of the electrolyte region can accept two $e^-$ and
decompose.  Electron motion of this nature may be sensitive to
details of DFT functionals (e.g., accuracy of predicted conduction band
positions).  Fortunately, such negative potentials vs.~Li$^+$/Li(s) are not
relevant to battery operations, where voltage control is exercised to
prevent over-charging.  Under normal conditions, the electrolyte
conduction band mininum resides above the anode Fermi level
(Fig.~\ref{fig6}a or~b).  Fluctuations in EC geometries 
are then required to lower EC lowest unoccupied molecular orbital
levels and permit $e^-$ transfer to EC.  Such geometric fluctuations help
surmount $e^-$ transfer barriers associated with reorganization free energies
enunciated in Marcus Theory.\cite{ald}

We have not observed Li(s) plating on LiC$_6$ surfaces even at
the lowest voltages because Li$^+$ diffusion and nucleation to form Li(s)
clusters occur on long timescales.

\subsection{EC/Li(s) Interface: Preliminary Investigations}
\label{li-interface}

\begin{figure}
\centerline{\hbox{ \epsfxsize=2.50in \epsfbox{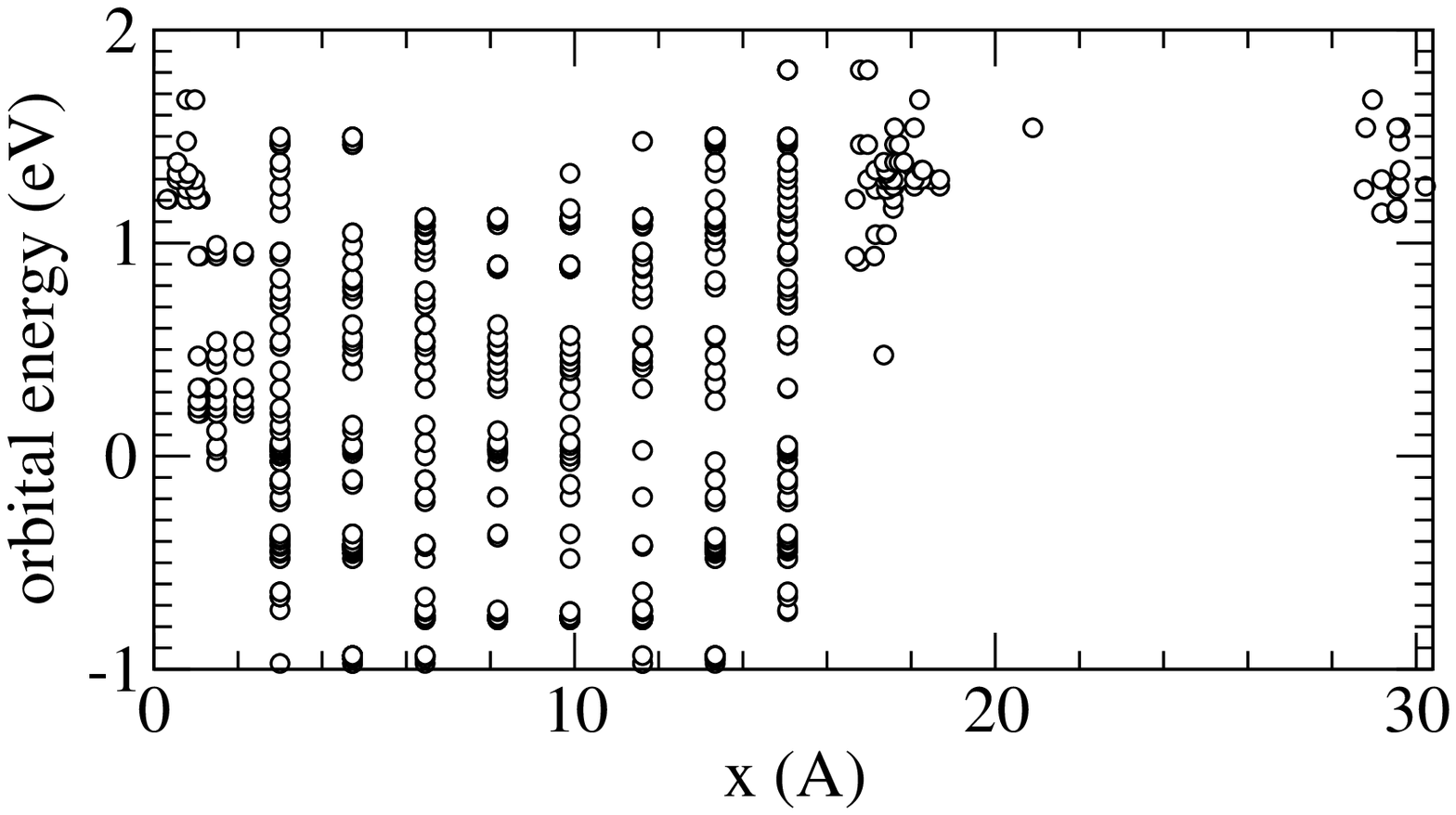} (a)}}
\centerline{\hbox{ \epsfxsize=2.50in \epsfbox{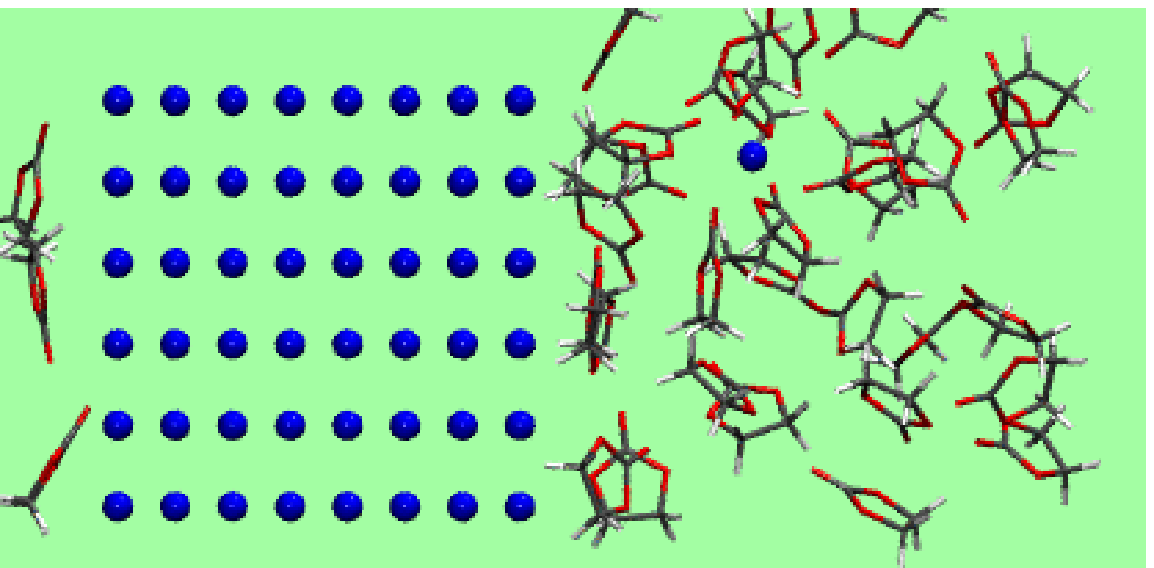} (b)}}
\caption[]
{\label{fig8} \noindent
(a) Instantaneous Kohn-Sham orbital decomposed on to atoms at their
$x$-coordinates.  $2<x<14$~\AA\, denotes the Li(s) region; outside
that range resides the electrolyte.  The Fermi level is at $E$=0.0~eV.
This simulation cell has a net $+|e|$ charge due to the solvated Li$^+$.
}
\end{figure}

It must be stressed that other interfaces may exhibit very different behavior.
For example, EC molecules directly coordinated to more reactive electrode
surfaces may more readily participate in
$e^-$ transfer due to strong electronic coupling.  We have re-analyzed
the initial configuration of an AIMD simulation of EC/Li(s) (100) interfaces
reported in Ref.~\onlinecite{ald}, in the absence of excess electrons
(i.e., at instantaneous lithiation potential of zero charge, Fig.~\ref{fig8}).
The electrolyte conduction band minimum lies at least 2.0~eV above the Li metal
Fermi level (not shown in Fig.~\ref{fig8}a). However, there is significant
hybridization between Fermi level metallic states on the electrodes and EC
orbitals spatially located near metal surfaces.  When an AIMD trajectory is
launched from this configuration, EC molecules on the surface rapidly
absorb $e^-$ and decompose to form mostly CO gas.\cite{ald}  Using a
hybrid DFT functional does not change this picture.  In contrast, on LiC$_6$
basal planes, the Kohn-Sham conduction bands in the electrolyte region
exhibit negligable coupling with the anode near the Fermi level unless
$\sigma$ is large and negative (Fig.~\ref{fig6}).

Explicit $\Delta G_t$ calculations are unlikely to succeed for pristine Li(s)
anodes because their extreme reactivity causes rapid solvent decomposition and
precludes sufficient AIMD sampling.  Instead, in Table~\ref{table3}, we
have estimated the LPZC by calculating contributions to the free energy of
Li$^+$ transfer from Li(s) to liquid EC if interfacial effects were absent
and the surface were uncharged.  Analogous to LiC$_6$, LPZC is large and
positive under these assumptions.  Yet, as mentioned above, EC rapidly
decomposes on the charge-neutral Li metal surface, contrary to EC on LiC$_6$
basal planes where reductive decomposition is observed only at large negative
potentials.  This comparison emphasizes that rates of $e^-$ transfer and
electrolyte decomposition strongly depend on surface details.

\section{Discussions}
\label{discuss}

So far we have performed voltage calculations on one type of electrode surface.
At the same applied voltage, different anode crystal facets should exhibit
variations in surface charge densities.  On carbon surfaces containing
edge sites -- e.g., pockets framed by C=O functional groups where Li$^+$ can
be strongly trapped\cite{pccp} -- the present formalism can determine the
instantaneous voltage with a fixed number of Li$^+$ trapped at edge sites if we
virtually transfer a Li$^+$ from the anode interior to the solvent.   At
longer timescales such that Li$^+$ can desorb from surface sites, the
potential becomes a function of both excess $e^-$ surface density and the amount
of bound Li$^+$ on the electrode surface in equilibrium with mobile Li$^+$ in
the liquid region.  The resulting surface concentration of Li$^+$ ions
(which are likely coordinated to some solvent molecules) may be 
substantially different from zero temperature DFT estimates performed
in the absence of the liquid electrolyte.\cite{interface,meng_review}
In a similar vein, on SEI-covered anodes, Li$^+$ (and perhaps less
likely, PF6$^-$) may adsorb on the SEI surface, and the total net charge
of a SEI-covered anode model may have a surface charge-potential relationship
quite different from that in Fig.~\ref{fig4}.  The goal in this work is to
use free energy methods to elucidate initial SEI formation, {\t prior to}
full SEI formation, but the ability to compute surface charge/voltage
profiles for SEI-coated electrodes will be essential for future studies of 
voltage-dependent Li$^+$ transport through the SEI film.

The spatial distribution of mobile Li$^+$ ions is not fully converged within
AIMD timescales currently available.  Hence it is difficult to quantify
the structure or function of the double layer in the simulations reported in
this work.  It is possible that the up to $\sim$0.12~eV statistical
uncertainties and the small Debye screening lengths have obscured double-layer
effects.  Advanced simulation techniques to accelerate salt sampling
will be an extremely valuable improvement.  However, we stress that
the initial AIMD configuration is pre-equilibrated by Monte Carlo 
simulations of classical force field models, where the electrodes are
not polarizable but exhibit the expected constant surface charge
density.  Therefore the electric double layer should be represented to the
extent that the force fields for electrodes and electrolytes are accurate.

Previous studies have shown that the M06-L functional yields the most
accurate Li$^+$ solvation free energy in acetonitrile.\cite{liox}
The M05-2X and PBE-derived functionals have also been shown to predict
Li$^+$/EC binding energies that differ by up to a few kcal/mol
($\sim$0.12~eV).  This can be a source of small systematic error in
$\Delta G_t$, which involves Li$^+$ solvation in EC liquid as an end point.

We have not considered the different stages of lithium-intercalated graphite.
At higher voltages, LiC$_{12}$, LiC$_{18}$, LiC$_{24}$ stochiometries,
and even graphite free of Li content, dominate.\cite{persson,holzwarth}  AIMD
simulations of interfaces are not
ideally suited to accommodating the changes in Li-content and lattice constant
variations as these stages transform into each other.  Instead we have focused
on the LiC$_6$ stoichiometry, consistent with Li intercalation
at the lowest potential, to facilitate the study of liquid-solid interfacial
effects.  We have also frozen all Li ions inside the slab so far.  Finally,
dispersion-corrected DFT can be used in the future.\cite{grimme}  This
is not expected to change the results substantially because AIMD simulations
start from classical force field-initiated configurations which are partly
determined by dispersion forces.

\section{Conclusions}
\label{conclude}

In this work, we have applied AIMD simulations to calculate the free energy
of Li$^+$ transfer ($\Delta G_t$) from an electronically conducting LiC$_6$
anode to liquid EC electrolyte, in condensed-phase settings appropriate
to lithium ion batteries.  We have correlated $\Delta G_t$ with the voltage
on the anode, which in turn depends on the net surface
charge.  Negative surface charge densities are compensated
by mobile, solvated Li$^+$ in the electrolyte in charge-neutral simulation
cells.  The approach, which does not require a vacuum gap, should be
rigorous for modeling electrochemical reactions on macroscopic metallic
electrodes in lithium ion batteries in the limit of large simulation cell
sizes and long trajectories.  Even in the present applications to
nanoscale simulation cells, the results are useful for calibrating 
voltages that cause low-barrier electrolyte decomposition reactions in
the same simulation cells.\cite{pccp,ald,mno,oakridge}  These calculations
can also potentially be corroborated with state-of-the-art nanoelectrochemical
measurements.

As a proof-of-principle example, we have considered LiC$_6$ basal planes
and their interfaces with ethylene carbonate (EC).  These electrochemically
inert graphite surfaces slow down EC decomposition and permit sufficient
sampling of $\Delta G_t$ over picosecond timescales.  Only at large negative
potentials relative to Li$^+$/Li(s) are electrolyte decomposition events
observed in picosecond timescales.  We predict that the basal planes need
to be negatively charged to retain Li$^+$.  A ``lithiation potential of
zero charge'' (LPZC) of 1.24~V vs.~Li$^+$/Li (s) is predicted for
LiC$_6$ basal planes.  This quantity is not measurable; Li de-intercalation
would have occurred at this voltage if edge planes were present to permit it.

At present, the statistical uncertainty in these computationally intensive
voltage calculations is on the order of 0.12~V.  The spatial
distributions of mobile Li$^+$ may not be extremely well-converged within
AIMD time scales used.  However, our empirical finding is that voltages
predicted with salt present do not exhibit significantly larger statistical
uncertainties than simulations conducted without mobile ions.  Our
qualitative conclusion about average surface charges at electrode/electrolyte
interfaces is robust, well within the margins of estimated  uncertainties.
Interesting future applications include anode surfaces which are more
reactive, e.g., graphite edge planes and electrode surfaces containing
spatial/chemical inhomogeneities.

\section*{Acknowledgement}

We thank John Sullivan, Michiel Sprik, Andrew Leenheer, Marie-Pierre Gaigeot,
Marialore Sulpizi, Oleg Borodin, Kevin Zavadil, and Peter Feibelman for
useful discussions.  Sandia National Laboratories is a multiprogram laboratory
managed and operated by Sandia Corporation, a wholly owned subsidiary of
Lockheed Martin Corporation, for the U.S.~Deparment of Energy's National
Nuclear Security Administration under contract DE-AC04-94AL85000.
AIMD simulations were supported by Nanostructures for Electrical
Energy Storage (NEES), an Energy Frontier Research Center funded by
the U.S.~Department of Energy, Office of Science, Office of Basic Energy
Sciences under Award Number DESC0001160.  Classical force field simulations
were funded by Sandia's Laboratory-Directed Research and Development program.

\section*{Supporting Information Available}

Further documentation are available regarding classical force field
calculations, snapshots and Li$^+$ distributions along AIMD trajectories,
and DFT work function predictions.  This information is available free
of charge via the Internet at {\tt http://pubs.acs.org/}.


\begin{references}
\bibitem{except}
Exceptions are calculations that include H$^+$ and compensating $e^-$
on metal electrodes.  See Ref.~\onlinecite{rossmeisl13}.
\end{references}
\end{document}